\documentclass[prd,preprint,superscriptaddress]{revtex4}
\usepackage{revsymb}
\usepackage{amsmath}
\usepackage{graphicx}
\usepackage{bm}
\newcommand{\be}{\begin{equation}}
\newcommand{\ee}{\end{equation}}
\newcommand{\ben}{\begin{eqnarray}}
\newcommand{\een}{\end{eqnarray}}

\begin{document}
\title{The generalized second law in phantom dominated universes in the presence of black holes}
\author{Germ\'{a}n Izquierdo\footnote{E-mail address: german.izquierdo@uab.es}}
\affiliation{Departamento de F\'{\i}sica, Facultad de Ciencias,
Universidad Aut\'{o}noma de Barcelona, 08193 Bellaterra
(Barcelona), Spain}
\author{Diego Pav\'{o}n\footnote{E-mail address: diego.pavon@uab.es}}
\affiliation{Departamento de F\'{\i}sica, Facultad de Ciencias,
Universidad Aut\'{o}noma de Barcelona, 08193 Bellaterra
(Barcelona), Spain}

%%%%%%%%%%%%%%%%%%%%%%%%%%%%%%%%%%%%%%%%%%%%%%%%%%%%%%%%%%%%%%%%%%%%%%%%%%%%%
\begin{abstract}
This Letter considers the generalized second law of gravitational
thermodynamics in two scenarios featuring a phantom dominated
expansion plus a black hole. The law is violated in both
scenarios.
\end{abstract}

\maketitle

%%%%%%%%%%%%%%%%%%%%%%%%%%%%%%%%%%%%%%%%%%%%%%%%%%%%%%%%%%%%%%%%%%%%%%%%%%%%%%

Current observational evidence leaves enough room for the dark
energy field driving the present accelerated expansion of the
Universe to be of phantom type \cite{evidence} -a form of energy
that  violates the dominant energy condition (DEC). If future
experiments come to confirm this, it will entail a profound impact
on cosmology and field theory. On the one hand, fields violating
the DEC (i.e., satisfying $\rho + p < 0$) face quantum
instabilities \cite{q-instabilities} and may drive, under certain
conditions, the scale and Hubble factors as well as the curvature
to diverge in a finite time ripping apart every bound structure,
from galaxy clusters down to atomic nuclei \cite{rip}. On the
other hand, as shown by Babichev {\em et al.} \cite{babichev},
black holes by accreting phantom energy lose mass and eventually
disappear altogether. This is easy to understand. In a general
accreting process the black hole mass rate is proportional to $A\,
(\rho + p)$, where $A$ is the area of the black hole horizon and
$\rho$ and $p$ are the energy and pressure of the accreted fluid,
respectively. If the latter fulfills the DEC, the black hole mass
will increase otherwise it will decrease. In this second case, the
immediate consequence is that the area of the black hole horizon
will go down along with its entropy, which is given by $
S_{BH}=4\pi \, M^{2}$ with $M$ the mass of the (Schwarschild)
black hole\footnote{We use units in which $c= G = \hbar = k_{B} =
1$.}.

A somewhat similar situation arises in the process of Hawking
radiation. There the decrease of the black hole mass  can be
traced to the accretion of virtual particles of negative energy.
However, as shown by Bekenstein \cite{jakob}, the overall entropy
does not diminish for the emitted radiation offsets for the loss
of black hole entropy. Nevertheless, in the case contemplated by
Babichev {\em et al.} there is no particle emission to make up for
the decrease of the black hole entropy. Thus, unless it gets
compensated by a corresponding increase in the area of the future
cosmic horizon and/or the phantom entropy, the generalized second
law (GSL) of gravitational thermodynamics will be violated. (The
latter, asserts that the entropy of matter and fields inside the
horizon plus the entropy of the horizon is a non-decreasing
quantity). The target of this Letter is to explore this.

Before going any further, it is expedient to recall that future
event horizon possess an entropy proportional to its area. In a
strict sense this has been proven rigourously just for the de
Sitter horizon \cite{gary}. However, it is only natural to
associate an entropy to the horizon area as it measures our lack
of knowledge about what is going on beyond it. This is why the
alluded proportionality is believed to hold true also in more
general space-times \cite{pollock,more-general}.

Babichev {\em et al.} consider a spatially flat
Friedmann--Robertson--Walker universe filled by a phantom fluid,
of energy density and pressure $\rho = - \frac{1}{2}\dot{\phi}^{2}+%
V(\phi)$ and $p = - \frac{1}{2}\dot{\phi}^{2} - V(\phi)$,
respectively, that dominates the expansion, and a Schwarzschild
black hole. If this is massive enough, its mass decrease via
Hawking radiation (proportional to $ M^{-2}$) can be safely
ignored and one can write $ \dot{M} = - 16 \pi \, M^{2}\,
\dot{\phi}^{2}$ regardless of the phantom potential. The latter is
felt on $\dot{M}$ through the field rate only.

In a recent paper \cite{plb}, we demonstrated that in a universe
dominated by phantom energy the GSL is satisfied at least in two
separate scenarios: $(i)$ When the equation of state parameter of
the fluid, $w \equiv p/\rho$, is a constant and, $(ii)$ when the
potential follows the power law $V(\phi) = V_{0}\, \phi^{\alpha}$
with $\alpha$ a constant parameter lying in the range $0< \alpha
\leq 4$ \cite{sami}. In both cases, it was found that the entropy
of the phantom fluid is negative\footnote{The negative character
of phantom's entropy was previously noted in Refs.
\cite{negative}.}, augments with the expansion of the universe and
that the entropy of the future cosmic horizon diminishes. The
latter can be written as $S_{H} =  {\cal A}/4 $, with ${\cal A} = 4 \pi\,%
R_{H}^{2}$ the area of the horizon. In general, the radius of the
future cosmic horizon, $R_{H}=a(t)\,\int_{t}^{\infty}{dt'/a(t')}$,
must be calculated via the scale factor of the metric which, in
its turn, is governed by the matter and fields filling the
universe.

\underline{In the first scenario} ($w =$ constant $< -1$, and
$V(\phi)\propto \exp(-\lambda \phi)$ with $\lambda = 4\,%
\sqrt{\pi/n}$), the corresponding scale factor obeys $a(t)%
\propto (t_{\ast} -t)^{-n}$ where $t \leq t_{*}$ and $0 <n = -
\frac{2}{3(1+w)}$, with $t_{\ast}$ being the big rip time
\cite{rip}. As a result, $R_{H} = (t_{\ast} -t)/n$ whence the
entropy of the cosmic horizon diminishes. Using Gibb's equation
and assuming thermal equilibrium between the phantom fluid and the
horizon it follows that the entropy of the former inside the
horizon obeys $S(t)= -S_{H}(t)$ -see Ref.\cite{plb} for details.
In consequence, the GSL is preserved but it also follows that if a
black hole is present, the GSL will no longer stand, i.e., we will
have $\dot{S} + \dot{S}_{H} + \dot{S}_{BH} < 0$ instead.
Obviously, the black hole must be small enough not to
significantly alter the scale factor quoted above for it was
calculated on the premise that the only energy source of the
Einstein field equations was the phantom field.

\underline{In the second scenario}, the phantom potential follows
the power law of above and the radius of the future cosmic horizon
is given by
\\
\begin{equation}
R_{H} = x^{\alpha/4}\, \Gamma \left(1- \frac{\alpha}{4}, x\right)
\, \frac{e^{x}}{H} \; \qquad \;  \left(x\equiv
\frac{4\pi}{\alpha}\phi^{2}\right) \, , \label{event2}
\end{equation}
\\
where
\\
\begin{equation}
H = \sqrt{\frac{2 V_{0}}{3}} \, (4%
\pi)^{\frac{1}{2}-\frac{\alpha}{4}} \alpha^{\alpha/4} \,
x^{\alpha/4}
\label{hfactor}
\end{equation}
\\
is the Hubble factor, $\dot{a}(t)/a(t)$, and $\Gamma \left(1 -
\frac{\alpha}{4}, x \right)$ the incomplete gamma function. As a
consequence, $R_{H}$ diminishes as the universe expands. In its
turn, the scalar field reads
\\
\ben
\phi(t)&=& \left[ \phi _{i}^{\frac{4-\alpha }{2}}+\sqrt{\frac{V_{0}}{24 \pi}}%
\, \frac{\left( 4-\alpha \right) \alpha }{2}\left( t-t_{i}\right) \right] ^{%
\frac{2}{4-\alpha }} \qquad \quad (0 < \alpha <4), \\
\phi(t)&=& \phi_{i} \, \exp\left[4 \sqrt{\frac{V_{0}}{24 \pi}}\,%
(t-t_{i})\right] \qquad (\alpha = 4).
 \label{field}
 \een
 \\
Note that in both cases $\dot{\phi} >0$. Here the $i$ subscript
indicates that the corresponding quantity is to be evaluated at
some suitable initial time, e.g. when the phantom energy begins to
dominate the expansion.

The time derivative of the horizon plus the phantom fluid can be
written as
\\
\begin{equation}
 \dot{S} + \dot{S}_{H}= \pi
\sqrt{\frac{3}{2V_{0}}}\, (4\pi)^{-\frac{1}{2}+\frac{\alpha}{4}}\,
\alpha^{-\alpha/4} \, R_{H}\, H \left[\Gamma \left( \frac{4-\alpha
}{4},\, x\right)\, e^{x} \left(2 + \frac{\alpha}{2x}\right)- 2%
x^{-\alpha/4} \right] ,\, \end{equation}
\\
which is a positive--definite quantity for any finite $x$.

From the definition of $x$ it is readily seen that
$\dot{\phi}=\sqrt{\alpha/(4\pi)}\, H/(2\sqrt{x})$, and in virtue
of this equation the black hole mass rate is
\\
\begin{equation}
\dot{M} = - \alpha \frac{H^{2}}{x}\, M^{2}\, ,
\label{Mdot}
\end{equation}
\\
thereby
\\
\begin{equation}
\dot{S}_{BH}=- 8\pi\alpha \frac{H^2}{x}\, M^{3}.
\label{dotSBH}
\end{equation}

The GSL, $ \dot{S}_{BH}+\dot{S} + \dot{S}_{H}\geq 0$, will be
satisfied provided the black hole mass does not exceed the
critical value
\\
\begin{eqnarray}
M_{cr}&=&
\sqrt{\frac{3}{8V_{0}}}(4\pi)^{-\frac{1}{2}+\,\frac{\alpha}{4}}\,
\alpha^{-\frac{1}{3}-\,\frac{\alpha}{4}}x^{\frac{1}{3}-\frac{\alpha}{12}}
\nonumber \\
&\times&\left\{e^{x}\Gamma \left(\frac{4-\alpha }{4},\,
x\right)\left[\Gamma \left( \frac{4-\alpha }{4},\, x\right)\,
e^{x} \left(2 + \frac{\alpha}{2x}\right)- 2 x^{-\alpha/4}
\right]\right\}^{\frac{1}{3}}\, .
\end{eqnarray}

As inspection reveals, $M_{cr}$ decreases with $\phi^{2}$ (and
therefore with time) at fixed $\alpha$. This stems from the fact
that $\dot{S} +\dot{S}_{H}$ is a decreasing function.

Again, the the black hole mass has to be low enough  to not
significantly modify the scale factor nor the cosmic horizon
radius, i.e., we must have $M \ll  E_{\phi}$, where $E_{\phi}=%
\frac{4\pi}{3}R_{H}^3 \, \rho$ is the energy of the phantom fluid
inside the horizon. In terms of $x$ it reads,
\\
\begin{equation}
E_{\phi}=\sqrt{\frac{3}{8V_{0}}}(4\pi)^{-\frac{1}{2}+\,\frac{\alpha}{4}}
\; \alpha^{- \frac{\alpha}{4}}\;
x^{\frac{\alpha}{2}}\left[e^x\,\Gamma \left( \frac{4-\alpha
}{4},\, x\right) \right]^3. \label{Ephi}
\end{equation}

To most direct way to see whether the GSL is violated or not is to
compare the evolution of $E_{\phi}$, $M_{cr}$, and $M$. The
expression for the latter follows from integrating Eq.
(\ref{Mdot}):
\\
\begin{equation}
M = \frac{M_{i}}{1 + M_{i} \,\sqrt{\frac{2V_{0}}{3}}%
(4\pi)^{\frac{1}{2}-\frac{\alpha}{4}}\; \alpha^{\alpha/4}\;
(x^{\alpha/4}-x_{i}^{\alpha/4} )} \; ,
\label{M(x)}
\end{equation}
\\
with $M_{i} \equiv M(x = x_{i})$.

Inspection of the last tree equations alongside their expressions
for $x \gg 1$, namely, $E_{\phi} \sim x^{-\alpha/4} + {\cal%
O}(x^{-\frac{\alpha}{4}-3})$, and $M_{cr} \sim
x^{-\frac{1}{3}-\frac{\alpha}{4}}+ {\cal
O}(x^{-\frac{\alpha}{4}-\frac{2}{3}})$, which follow from the
relation \cite{Handbook}
\\
\[
e^{x}\, \Gamma \left(1-\frac{\alpha}{4}, x\right) \sim
x^{-\alpha/4} \left( 1 -\frac{\alpha}{4x}+ \frac{\alpha^{2} + 4
\alpha}{16 x^{2}} + .... \right) \, ,
\]
\\
reveals that for reasonable values of $M_{i}$, initially (i.e.,
for $x_{i}\sim {\cal O}(1)$) we will simultaneously have $M <%
M_{cr}$ and $M \ll E_{\phi}$, and somewhat later we will have
$M_{cr} < M \ll E_{\phi}$, instead. The latter corresponds to a
violation of the GSL.

\begin{figure}[tbp]
\includegraphics*[scale=0.4,angle=0]{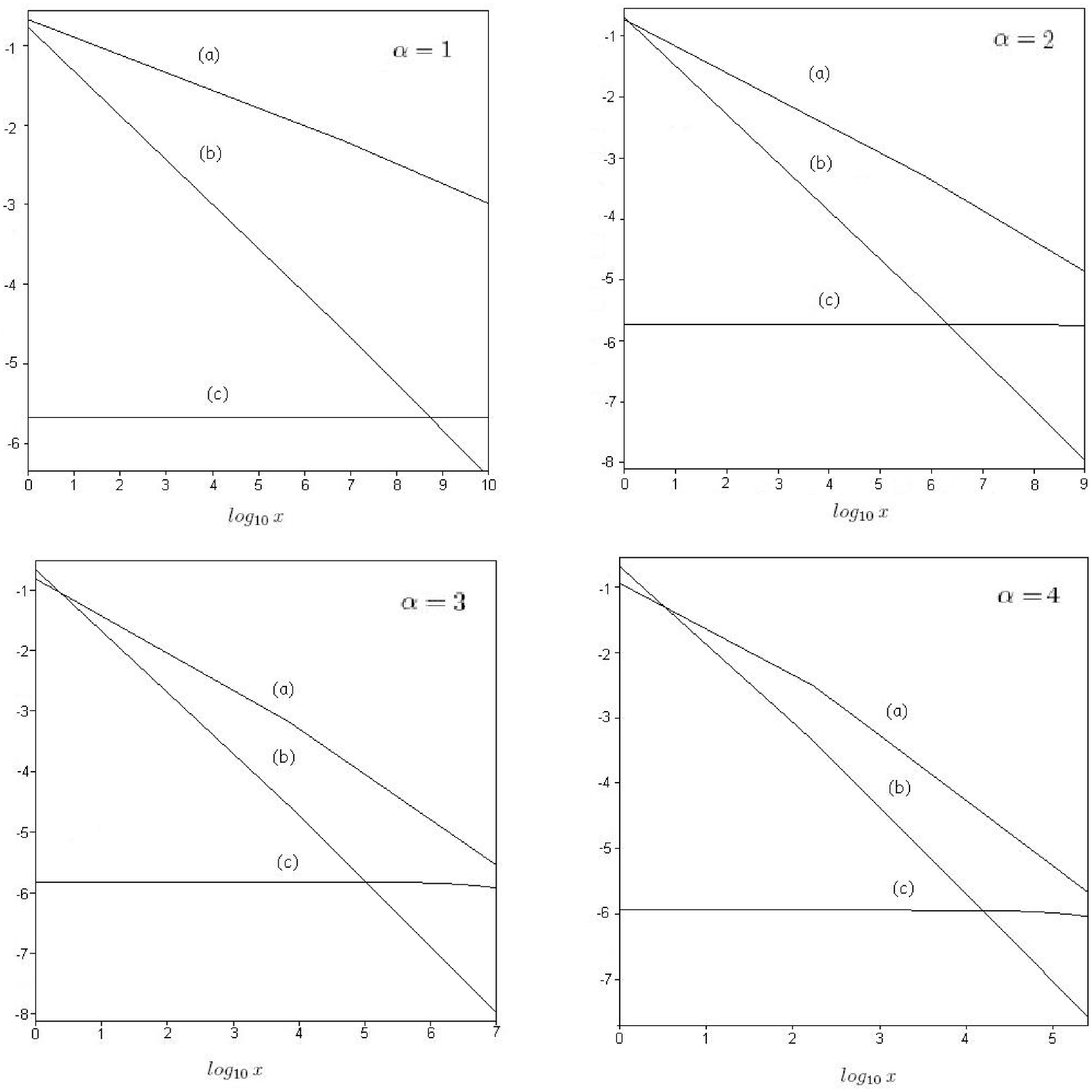}
\caption{Evolution of  $E_{\phi}$ (a),  $M_{cr}$ (b), and $M$ (c),
{\em vs} $x$ in logarithmic scales. The initial mass of the black
hole was chosen $M_i=10^{-5}E_{\phi}(x_i)$. As it can be
appreciated in the four panels, the black hole mass exhibits a
much steeper decrease than the critical mass. Consequently,
irrespective of the $M_{i}$ value, sooner or later we will have
that $M \geq M_{cr}$ and thus the GSL will fail. Later on, $M$
will become non-negligible (compared to $E_{\phi}$) and our model
will no be longer  valid. In plotting the graphs we have set
$V_{0}^{-1/2}= 1$, and $x_{i}=1$, for simplicity.}
\label{fig:alpha}
\end{figure}

Indeed, as Fig. \ref{fig:alpha} shows, for every $\alpha$ there is
an initial $x$ interval in which $M_{cr}$ is larger than $M$. The
GSL is fulfilled in that interval. However, later on, $M_{cr}$
becomes smaller that $M$ (with $M \ll E_{\phi}$) and  the GSL is
violated. Still further ahead, $M$ is no longer negligible against
$E_{\phi}$. Thus, for sufficiently large $x$, this assumption
breaks down and from there on it cannot be said whether the GSL is
violated or not. At any rate, for any $\alpha$, there is an ample
$x$ interval in which it can be safely said that the GSL does not
hold.

Interestingly enough, while in the second scenario, which is big
rip free, there is enough room for the GSL to be fulfilled there
is no room whatsoever in the first scenario, which features a big
rip.

In the light  of the foregoing results some reactions may arise:
$(a)$ Some phantom energy fields might be physical but not those
considered in this Letter. In fact, some predictions lending
support to phantom fields may have come from an erroneous
interpretation of the observational data \cite{das}. $(b)$ The GSL
was initially formulated for systems fulfilling the DEC, so there
is no reason why it ought to be satisfied for systems that violate
it. What is more, in a strict sense, a general proof of the GSL
even for systems complying with the DEC is still lacking
\cite{rmw}, therefore there should be no wonder that it fails in
some specific instances.

It is for the reader to decide which of these views, if any, is
more to his/her liking.

Yet, one may argue that it is unclear that black holes retain
their thermodynamic properties (entropy and temperature) in
presence of a field that does not comply with the DEC. In such an
instance, one may think, that there is no room for the black hole
entropy in the expression for the GSL. However, the latter is
often formulated by replacing $S_{BH}$ by the black hole area.
Again, this variant of the GSL will fail in the two cases of
above. In all, if eventually phantom energy is proven to be a
physical reality, it will mean a serious threat to the generalized
second law.

\acknowledgments{The authors are grateful to David Coule for
correspondence on the subject of this Letter. This work was
partially supported by the old Spanish Ministry of Science and
Technology under Grant BFM--2003--06033, and the ``Direcci\'{o}
General de Recerca de Catalunya" under Grant  2005 SGR 000 87.}

\end{document}